\documentclass[conference]{IEEEtran}
\usepackage{cite}
\usepackage{amsmath,amssymb,amsfonts}
\usepackage{algorithmic}
\usepackage{graphicx}
\usepackage{textcomp}
\usepackage{xcolor}
\usepackage{array}
\newcolumntype{P}[1]{>{\centering\arraybackslash}p{#1}}

\usepackage{makecell}
\usepackage{hhline}\usepackage{booktabs}
\usepackage{comment}

\usepackage{stackengine}

\usepackage{siunitx}
\usepackage{multirow}

\def\BibTeX{{\rm B\kern-.05em{\sc i\kern-.025em b}\kern-.08em
    T\kern-.1667em\lower.7ex\hbox{E}\kern-.125emX}}

\begin{document}

\title{Detection of Pedestrian Turning Motions to Enhance Indoor Map Matching Performance}

\makeatletter
\newcommand{\linebreakand}{%
  \end{@IEEEauthorhalign}
  \hfill\mbox{}\par
  \mbox{}\hfill\begin{@IEEEauthorhalign}
}
\makeatother

\author{\IEEEauthorblockN{Seunghyeon Park}
\IEEEauthorblockA{\textit{School of Integrated Technology} \\
\textit{Yonsei University}\\
Incheon, Korea \\
seunghyeon.park@yonsei.ac.kr} 
\and
\IEEEauthorblockN{Taewon Kang}
\IEEEauthorblockA{\textit{School of Integrated Technology} \\
\textit{Yonsei University}\\
Incheon, Korea \\
taewon.kang@yonsei.ac.kr}
\and
\IEEEauthorblockN{Seungjae Lee}
\IEEEauthorblockA{\textit{School of Integrated Technology} \\
\textit{Yonsei University}\\
Incheon, Korea \\
seungjae@yonsei.ac.kr}
\linebreakand
\IEEEauthorblockN{Joon Hyo Rhee${}^{*}$} 
\IEEEauthorblockA{\textit{Korea Research Institute of Standards and Science} \\
Daejeon, Korea \\
jh.rhee@kriss.re.kr}
{\small${}^{*}$ Corresponding author}
}

\maketitle

\begin{abstract}
A pedestrian navigation system (PNS) in indoor environments, where global navigation satellite system (GNSS) signal access is difficult, is necessary, particularly for search and rescue (SAR) operations in large buildings. 
This paper focuses on studying pedestrian walking behaviors to enhance the performance of indoor pedestrian dead reckoning (PDR) and map matching techniques. 
Specifically, our research aims to detect pedestrian turning motions using smartphone inertial measurement unit (IMU) information in a given PDR trajectory.
To improve existing methods, including the threshold-based turn detection method, hidden Markov model (HMM)-based turn detection method, and pruned exact linear time (PELT) algorithm-based turn detection method, we propose enhanced algorithms that better detect pedestrian turning motions.
During field tests, using the threshold-based method, we observed a missed detection rate of 20.35\% and a false alarm rate of 7.65\%. 
The PELT-based method achieved a significant improvement with a missed detection rate of 8.93\% and a false alarm rate of 6.97\%. 
However, the best results were obtained using the HMM-based method, which demonstrated a missed detection rate of 5.14\% and a false alarm rate of 2.00\%.
In summary, our research contributes to the development of a more accurate and reliable pedestrian navigation system by leveraging smartphone IMU data and advanced algorithms for turn detection in indoor environments.
\end{abstract}


\section{Introduction}
A pedestrian navigation system (PNS) in indoor environments \cite{Kang21:Indoor, Park21:Indoor}, where global navigation satellite system (GNSS) \cite{Kim14:Comprehensive, Lee22:Optimal, DeLorenzo10:WAAS/L5, Park21:Single, Chen11:Real, Lee23:Seamless, Kim23:Machine, Kim23:Low} signal access is difficult, is necessary, particularly for search and rescue (SAR) \cite{Lee22:Evaluation, Lee23:Performance_Evaluation, Lee23:Performance_Comparison} operations in large buildings. 
Some systems rely on preinstalled infrastructure. 
These include pseudolites \cite{Angermann}, radio-frequency identifications (RFIDs) \cite{Fan17, Qi15} and ultra-wideband (UWB) radars \cite{Shin17:Autonomous}. 
However, these infrastructure installation and maintenance procedures require significant time and effort \cite{Lee22:Urban, Jeong20:RSS}. 
To navigate without the use of infrastructure, pedestrian dead reckoning (PDR) systems sometimes use an inertial measurement unit (IMU) mounted on the foot because the foot-mounted IMU has an explicit stance stage that can be used to suppress the long-term drift of inertial sensors using the zero-velocity update (ZUPT) extended Kalman filter (EKF) method \cite{Godha}.

Our work aims to investigate and leverage pedestrian walking behaviors to enhance the performance of indoor PDR and map matching techniques. 
PDR enables the tracking of a pedestrian's trajectory without relying on external signals or infrastructure. 
However, PDR is susceptible to cumulative errors over time, leading to a significant degradation in position estimation accuracy. 
By reliably detecting pedestrian turning motions within a given indoor PDR trajectory, we can identify the locations where turns occur, which can be considered as intersections of corridors. 
These intersections serve as reference points to improve the accuracy of indoor map matching techniques. 
Thus, our study focuses on the detection of pedestrian turning motions in the PDR trajectory.

Several methods to detect pedestrian turns have been developed by various researchers. 
Three representative methods to detect a pedestrian's turning motion are as follows. 
\begin{enumerate}
\item Threshold-based turn detection method \cite{Pham2017}
\item Hidden Markov model (HMM)-based turn detection method \cite{Zhou}
\item Pruned exact linear time (PELT) algorithm-based turn detection method \cite{Gu15:Foot}
\end{enumerate}
However, these methods have limited accuracy. 
There are turns that cannot be detected by these algorithms, and there are cases in which a turn is detected even when it is not a turn. 
Therefore, we propose methods to improve each of the three turn detection methods to enhance the performance.

\section{Threshold-Based Turn Detection Method} 
\label{sec:Threshold-BasedTurnDetectionMethod}

\subsection{Existing Threshold-Based Turn Detection Method}

The existing threshold-based method detects the turn if the changed value of the yaw angle between adjacent steps exceeds a predefined threshold. 
This method was developed in several studies \cite{Pham2017, Gal2020, Novak201418800} owing to its simplicity and cost-effectiveness. 
In the example of Fig. \ref{fig:hmm_map}, the pedestrian walked through an arbitrary indoor route while holding a smartphone, recording motion data of the pedestrian in a time series format using built-in IMU sensors. 
Then, the data was processed by a conventional PDR algorithm to plot the trajectory of the pedestrian. 
The dot (black and red) forming the trajectory in Fig. \ref{fig:hmm_map} shows the position estimate of each step. 
The black dots show the steps where the pedestrian walked straight, and the red dots show the steps where the pedestrian turned. 

However, this method shows a low detection accuracy when the length of a turn is longer than usual turns. 
If the turn consists of many steps, the yaw angle changes between adjacent steps in the same turn can vary.
Thus, some steps with yaw angle changes exceeding the threshold are declared as turns and the others are not, even though those steps belong to the same turn. 
As a result, a single turn can be recognized as multiple erroneous turns if this turn-detection method is applied. 

\begin{figure} 
    \centering 
    \includegraphics[width=0.6\linewidth]{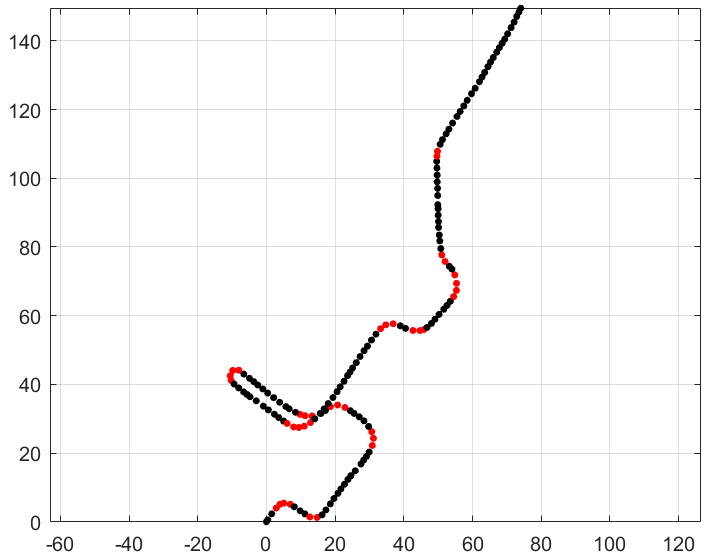}
    \caption{Map showing pedestrian steps measured by a smartphone IMU and processed by a conventional PDR algortihm.} \label{fig:hmm_map}
\end{figure}

\subsection{Proposed Threshold-Based Turn Detection Method}
\label{sec:ProposedThreshold}

To address the aforementioned weakness of the threshold-based turn detection method, we propose an additional algorithm. 
When a detected turn step is adjacent to previously detected turn steps, it is reasonable to consider it as part of the existing turn sequence. 
Therefore, in our algorithm, when multiple turns are in close proximity to each other, they are treated as a single turn. 
This approach effectively reduces the number of erroneously detected turns.
The performances of the existing and proposed methods are compared in Section \ref{sec:PerformanceComparison}.

\section{Hidden Markov Model (HMM)-Based Turn Detection Method}
\label{sec:HMM-BasedTurnDetectionMethod}

\subsection{Existing HMM-Based Turn Detection Method}

\begin{figure} 
    \centering 
    \includegraphics[width=0.85\linewidth]{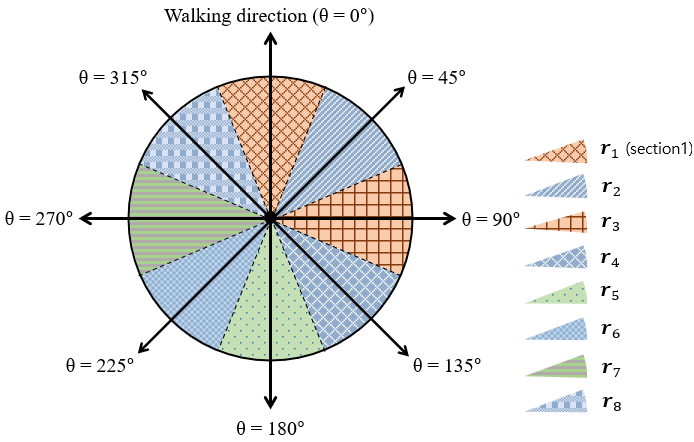}
    \caption{Pedestrian walking directions and regions.} \label{fig:hmm_direction}
\end{figure}

\begin{figure} 
    \centering 
    \includegraphics[width=0.85\linewidth]{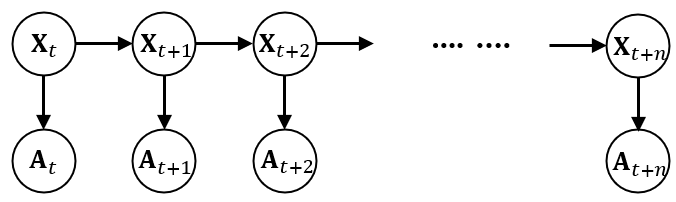}
    \caption{HMM model used for turn detection.} 
    \label{fig:hmm_model}
\end{figure}

The hidden Markov model (HMM) is used to deduce input information by statistically analyzing large amounts of output data \cite{Johansson:Bayesian, Turhan}. 
In order for HMM to be employed in modeling the pedestrian's walking behavior, the pedestrian's prior step information needs to be maintained to some extent.
Therefore, Gu \textit{et al.} \cite{Gu15:Foot} utilized the HMM algorithm by combining three walking steps as a single state of HMM assuming that a single turning motion consists of three walking steps. 
In this manner, this method can analyze the relationship between the six adjacent steps (i.e., two states of HMM), rather than two adjacent steps, while using the HMM algorithm for detecting a turning motion.

Specifically, the existing method \cite{Gu15:Foot} classifies the walking direction into eight regions according to the angle of the direction, as shown in Fig. \ref{fig:hmm_direction}.
The input used in this classification process is the walking direction angle $\theta$ for each step, and then the steps are classified into region $r$ to which $\theta$ belongs.
Since there are many possible values of a state with three walking steps (i.e., $8\times8\times8=256$ possible values), values that do not appear and values that appear at low frequency are removed, and only 14 values (i.e., $v_1, \ldots, v_{14}$) are configured through clustering in this approach. 

This method \cite{Gu15:Foot} does not detect or calculate the probability of the pedestrian turning on the following step; instead, it predicts the probability of being in each of the 14 values (i.e., $v_1, \ldots, v_{14}$). 
In Fig. \ref{fig:hmm_model}, the observation matrix $A_t$ contains ${P(v_1), P(v_2), \ldots, P(v_{14})}$, which represents the probabilities of each value that the state $X_{t+1}$ can have after the state $X_t$.

Therefore, to perform turn detection based on the information obtained from the existing HMM technique \cite{Gu15:Foot}, we utilized the probabilities of the possible values for the next state $X_{t+1}$, given the value of the current state $X_t$. 
For instance, suppose $X_t$ has the value $v_1$ and ${P(v_1)=0.6, P(v_2)=0.01, \ldots, P(v_{14})=0.1}$. 
In this scenario, $X_{t+1}$ has a 60\% probability of having the same value as $X_t$ due to $P(v_1)=0.6$. 
If the probability $P(v_t)$ of maintaining the current value $v_t$ of state $X_t$ is lower than 50\%, it is determined as a turn since the state value is likely to change with a probability greater than 50\%. 
Conversely, if the probability of maintaining the current state value is higher than 50\%, it is determined that the pedestrian does not make a turn for that state.

However, as depicted in Fig. \ref{fig:hmm_map}, the process of turning or walking straight does not always occur in three steps. 
There are instances where a turn is completed in one or two steps, and in some cases, a turn may take more than six steps. 
Consequently, the existing HMM approach exhibits limited accuracy in turn detection. 
To achieve more effective modeling, it becomes necessary to capture the broader human gait behavior. 
In this regard, we propose a modified HMM algorithm, which will be elaborated upon in the subsequent subsection.

\subsection{Proposed HMM-Based Turn Detection Method}
\label{sec:ProposedHMM}

In the proposed HMM-based turn detection method, we group steps exhibiting similar behavior into a new entity. 
This grouping is necessary because the number of steps required for a single turn is not always fixed, as assumed in \cite{Gu15:Foot}, where three steps were considered. 
In this paper, we refer to this linked group of steps as a ``block.'' 
Consequently, a block can consist of as few as one step or as many as ten steps, for example. 
In our HMM, each individual state corresponds to a single block. 
To determine the length of each block, we employed the Butterworth bandpass filter. 
This filter identifies the peak locations of yaw angle measurements, which can be considered as change points in a pedestrian's walking behavior.
The time interval between these peaks is then selected as the length of a block. 
This approach enhances the accuracy of comprehending pedestrian turning behavior in comparison to existing methods.

The main feature of a pedestrian's trajectory is the direction angle $\theta$.
This angle, $\theta$, is referred to as the difference between the body's yaw plane velocity vector and orientation vector.
We computed the orientation vector using the smartphone IMU's gyroscope measurements and obtained the velocity vector using its acceleration measurements.
We categorized $\theta$ into eight discrete state values based on the corresponding regions for each step, as shown in Fig. \ref{fig:hmm_direction}.
Therefore, a pedestrian's movement is considered a turn only when it deviates more than $22.5^\circ$ from the current direction.

The transition matrix and emission matrix, which are modeling parameters of the HMM, are constructed using the eight possible state values.
The process of constructing the parameters of HMM is as follows:

First, we obtain the region number for each block.
There are eight regions denoted as $\{r_1, r_2, r_3, \ldots, r_8\}$ as shown in Fig. \ref{fig:hmm_direction}. 
This region number is directly determined based on the $\theta$ information obtained earlier.
For example, if $22.5^\circ  \le \theta <  67.5^\circ $, the region number for that state or block is $r_2$.
The region of the pedestrian's first block is designated as $r_1$ and serves as the baseline for the entire walk.
If the pedestrian turns 30 degrees clockwise from the initial walking direction, for example, the region number changes from $r_1$ to $r_2$.

Secondly, we calculate the transition probabilities.
In the Markov chain, there are several states, and the transition probability refers to the probability of moving from one state to another \cite{Ye, Guogang}. 
Therefore, we calculated the probability of transitioning from one of the eight regions $\{r_1, r_2, r_3, \ldots, r_8\}$ to another, as presented in Table \ref{tab:transition}.

The third step is to obtain the emission probabilities, which represent the probability distribution of observing evidence $A_t$ in a given state $X_t$. 
Each state $X_t$ corresponds to one of the eight regions shown in Fig. \ref{fig:hmm_direction}. 
The evidence $A_t$ refers to the status of the pedestrian, indicating whether they are turning or walking straight. 
Consequently, the emission matrix represents the probabilities of the pedestrian turning or not turning at each region, as listed in Table \ref{tab:emission}.

A turn is detected if the probability of staying in the current state value at the following state is less than 0.5. 
The performance of both the existing and proposed methods is compared in Section \ref{sec:PerformanceComparison}.

\begin{table*}
\centering
\caption{Transition matrix used in the proposed HMM-based method \label{tab:transition}}
\begin{center}
\begin{tabular}{ |P{0.8cm}|P{0.9cm}|P{0.9cm}|P{0.9cm}|P{0.9cm}|P{0.9cm}|P{0.9cm}|P{0.9cm}|P{0.9cm}| }
 \hline
 \multicolumn{9}{|c|}{Transition Matrix} \\
 \hline
 region & $r_1$ & $r_2$ & $r_3$ & $r_4$ & $r_5$ & $r_6$ & $r_7$ & $r_8$\\
 \hline
 $r_1$   & 0.5921 &0.0542 &  0.0244 & 0.0105 & 0.0294 & 0.0114 & 0.0162 & 0.0705\\
 $r_2$   & 0.0950 &0.7808 &  0.1389 & 0.0056 & 0.0037 & 0.0030 & 0.0015 & 0.0041\\
 $r_3$   & 0.0257 &0.1220 &  0.7311 & 0.0670 & 0.0074 & 0.0000 & 0.0015 & 0.0062\\
 $r_4$   & 0.0515 &0.0079 &  0.0944 & 0.8406 & 0.1532 & 0.0068 & 0.0015 & 0.0041\\
 $r_5$   & 0.0277 &0.0068 &  0.0044 & 0.0664 & 0.6949 & 0.0781 & 0.0103 & 0.0073\\
 $r_6$   & 0.0416 &0.0090 &  0.0011 & 0.0050 & 0.1029 & 0.8225 & 0.1521 & 0.0093\\
 $r_7$   & 0.0238 &0.0056 &  0.0000 & 0.0019 & 0.0049 & 0.0698 & 0.7061 & 0.0870\\
 $r_8$   & 0.1426 &0.0102 &  0.0056 & 0.0031 & 0.0037 & 0.0083 & 0.1108 & 0.8104\\
 \hline
\end{tabular}
\end{center}
\end{table*}

\begin{table*}
\centering
\caption{Emission matrix used in the proposed HMM-based method \label{tab:emission}}
\begin{center}
\begin{tabular}{ |P{0.8cm}|P{0.9cm}|P{0.9cm}|P{0.9cm}|P{0.9cm}|P{0.9cm}|P{0.9cm}|P{0.9cm}|P{0.9cm}| }
 \hline
 \multicolumn{9}{|c|}{Emission Matrix} \\
 \hline
 turn & $r_1$ & $r_2$ & $r_3$ & $r_4$ & $r_5$ & $r_6$ & $r_7$ & $r_8$\\
 \hline
 O   & 0.0812 &0.1085 &  0.1300 & 0.1011 & 0.1360 & 0.0979 & 0.1773 & 0.1192\\
 X   & 0.9188 &0.8915 &  0.8700 & 0.8989 & 0.8640 & 0.9021 & 0.8227 & 0.8808\\
 \hline
\end{tabular}
\end{center}
\end{table*}

\section{Pruned Exact Linear Time (PELT) Algorithm-Based Turn Detection Method}
\label{sec:PELT-BasedTurnDetectionMethod}

\subsection{Existing PELT-Based Turn Detection Method}

To detect a turn, it is necessary to define the point at which the turn occurs, indicating a rapid change in the statistical properties \cite{Zhou, Killick}. 
In the pruned exact linear time (PELT) approach, pruning is carried out at the current time step if the current cost is less than the cost at the potential change-point plus the extra segment cost.
For a given set of data $X = (x_1,x_2,...,x_n)$, let $t_0=0$, $t_{k+1}=n$, and $t_1,t_2,...,t_k$ be the points of change in chronological order. 
These points of change are the ones that minimize the following equation \cite{Zhou}:
\begin{equation} 
    \begin{split}
         \sum_{i=1}^{k+1} & \left( C\left(x_{\mathrm{t_{\mathrm{i-1}}+1} : t_{\mathrm{i}}} \right) + \beta \right) 
    \end{split}
\end{equation}
where $x_{a:b}$ = ($x_a, x_{a+1},...,x_{b}$) ($a \le b$), $C(x_{a:b})$ is the cost function of $x_{a:b}$, and $\beta$ represents the penalization to prevent overfitting.

However, in the PELT approach, there were instances where a genuine turn was missed, occurring in 44.67\% of the cases.
To enhance the performance of the model, an anomaly detection algorithm was incorporated to improve its capabilities. 
Anomaly detection involves identifying rare observations that significantly deviate from the majority of the data and do not conform to normal behavior.
By incorporating this algorithm, the model becomes more adept at detecting such anomalies and improving overall performance.

\subsection{Existing IF-Based Turn Detection Method}

In our study, the isolation forest (IF) algorithm was utilized for anomaly detection.
The isolation forest operates on a tree-based structure by randomly partitioning the data to isolate all observations, as shown in Fig.\ref{fig:if_tree} \cite{Lee:Anomaly, Zhang:Isolation, Cheng:Outlier}. 
The algorithm is based on the simple principle that isolating anomalies is easier than isolating normal data \cite{Wang:research, Zhang:Data}. 
Outliers tend to be closer to the root node of the tree, resulting in shorter path lengths  \cite{Chabchoub, Raumonen}. 
By employing a collection of trees that are randomly fitted to the data, we can calculate the average depth at which outliers occur within the trees.
This average depth is then used to generate a final score reflecting the degree of ``outlierness'' for each data point.

By utilizing the IF algorithm, outliers can be effectively detected. 
We considered nine types of IMU data, including acceleration, gravitational force, and magnetic field along the x, y, and z axes, as the input data. 
It was observed that there are differences in this data within steps that involve turning \cite{Zhang23}. 
Therefore, turns can be detected through the IF algorithm, which detects outliers from the given data.

\subsection{Proposed PELT+IF-Based Turn Detection Method}
\label{sec:ProposedPELT}

\begin{figure} 
    \centering 
    \includegraphics[width=1.0\linewidth]{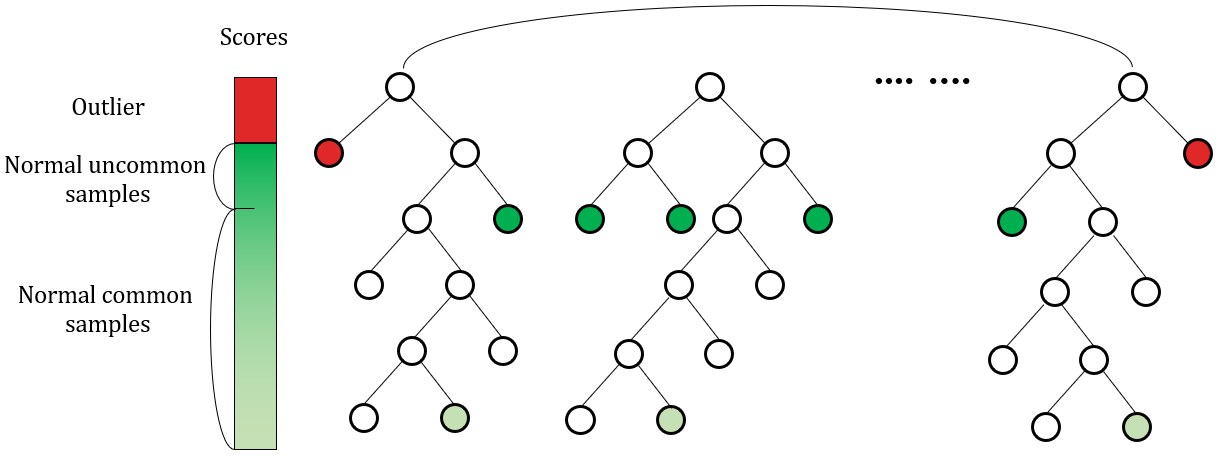}
    \caption{Tree of isolation forest.} 
    \label{fig:if_tree}
\end{figure}

We employed the IF algorithm to address the limitations of the PELT approach.
When using the PELT algorithm, the false alarm rate was low (6.97\%), but there was a significant missed detection rate (44.67\%). 
In cases where an outlier exhibited distinct characteristics, it was challenging for the PELT algorithm to recognize it as a ``point of change.''
However, it is highly probable that such outliers represent actual points of change, thus necessitating their detection through the use of the IF algorithm.

In the IF algorithm, a scoring process is employed to determine the extent to which a separated point should be considered an outlier. 
This scoring process helps establish a threshold for identifying outliers based on their deviation from the original data. 
The scoring calculation in the IF algorithm is performed using the following equation \cite{Zhang:Isolation}:
\begin{equation} 
    s\left(x,n\right) = 2^{-\frac{E(h(x))}{c(n)}} 
\end{equation}
Here, $h(x)$ represents the path length of the corresponding observation, $E(h(x))$ denotes the average path length for that observation across all iterations, and $c(n)$ represents the average path length of the tree used to determine $h(x)$.

For normal data, where the observed value $x$ is similar to the average path length, $E(h(x))$ converges to $c(n)$, resulting in a score of approximately 0.5 \cite{Sun, Xu}. 
However, if the observed value $x$ is an outlier, $E(h(x))$ tends to converge to 0, leading to a score close to 1. 
Since the maximum path length of the observation $x$ is $E(h(x)) = n-1$, the score $s(x,n)$ falls within the range of 0 to 1. 
A score greater than 0.5 indicates a higher likelihood of being an outlier, while a score less than or equal to 0.5 indicates normal data. 
The pollution parameter in IF \cite{Kim:study} controls the threshold for determining when a scored data point should be considered an outlier. 
We experimentally set the pollution parameter to 0.05 to achieve the best result.

\section{Test Results}
\label{sec:TestResults}

\subsection{Data Collection}
\label{sec:DataCollection}

A total of 58 instances of sensor log data collected in the laboratory were utilized as both input and output data values.
These nine types of IMU data encompassed the acceleration along the \textit{x}-axis, acceleration along the \textit{y}-axis, acceleration along the \textit{z}-axis, gravitational force along the \textit{x}-axis, gravitational force along the \textit{y}-axis, gravitational force along the \textit{z}-axis, magnetic field along the \textit{x}-axis, magnetic field along the \textit{y}-axis, and magnetic field along the \textit{z}-axis.

\subsection{Experiment Results}
\label{sec:PerformanceComparison}

The performance of the proposed methods in Sections \ref{sec:ProposedThreshold}, \ref{sec:ProposedHMM}, and \ref{sec:ProposedPELT} was evaluated by analyzing missed detections and false alarms.
Missed detection refers to cases where a turn is not recognized among all the actual turns.
On the other hand, false alarms occur when a turn is incorrectly identified even though no rotation has taken place.
The occurrence of these cases was quantified as probabilities.

\subsubsection{Threshold-Based Turn Detection Method Performance}

Among the 159 collected ambulation data, we computed the missed detection and false alarm rates for accurately detecting 1,979 executed turns. 
 In the existing threshold-based turn detection method, the missed detection rate was determined to be 20.35\%, while the false alarm rate was 10.45\%, as indicated in Table \ref{tab:threshold}. 
For the proposed threshold-based turn detection method, the missed detection rate was found to be 20.35\%, and the false alarm rate was 7.65\%.

\begin{table} 
\centering
\caption{Performance comparison between two threshold-based algorithms in turn detection}
\label{tab:threshold}
\begin{center}
{\renewcommand{\arraystretch}{1.4}
 \begin{tabular}[c]{>{\centering\arraybackslash}m{0.1cm}>{\centering\arraybackslash}m{1.3cm}>{\centering\arraybackslash}m{2.0cm} >{\centering\arraybackslash}m{2.0cm}>{\centering\arraybackslash}m{2.0cm}}
 \toprule
    {} & {} & \thead{Missed detection} 
       & \thead{False alarm} \\
    \noalign{\vspace{0pt}}
\hline
 & Existing algorithm & 20.35\% & 10.45\%  \\
 \noalign{\vspace{1pt}}
 \midrule
 & Proposed algorithm & 20.35\% & 7.65\% \\
 \noalign{\vspace{1pt}}
 \bottomrule
\end{tabular}}
\end{center}
\end{table}

\subsubsection{HMM-Based Turn Detection Method Performance}

The performance of the HMM methods is presented in  Table \ref{tab:hmm}.
When using the existing HMM algorithm, the missed detection rate was 7.11\%, and the false alarm rate was 30.33\%. 
This algorithm is considered unreliable as it frequently identifies steps as turns, even when no actual turns are performed.
In contrast, the proposed HMM algorithm achieved a missed detection rate of 5.14\%, and a false alarm rate of 2.00\%, which is significantly lower than the existing method. 
Particularly, there was a significant improvement in the false alarm rate by the proposed method.

\begin{table} 
\centering
\caption{Performance comparison between two HMM-based algorithms in turn detection}
\label{tab:hmm}
\begin{center}
{\renewcommand{\arraystretch}{1.4}
 \begin{tabular}[c]{>{\centering\arraybackslash}m{0.1cm}>{\centering\arraybackslash}m{1.3cm}>{\centering\arraybackslash}m{2.0cm} >{\centering\arraybackslash}m{2.0cm}>{\centering\arraybackslash}m{2.0cm}}
 \toprule
    {} & {} & \thead{Missed detection} 
       & \thead{False alarm} \\
    \noalign{\vspace{0pt}}
\hline
 & Existing algorithm & 7.11\% & 30.33\%  \\
 \noalign{\vspace{1pt}}
 \midrule
 & Proposed algorithm & 5.14\% & 2.00\% \\
 \noalign{\vspace{1pt}}
 \bottomrule
\end{tabular}}
\end{center}
\end{table}

\subsubsection{PELT+IF-Based Turn Detection Method Performance}

The missed detection rate accounted for 44.67\% of the total 1,979 turns, while the false alarm rate was 6.97\%.
In the case of the PELT-based turn detection method, a noticeable ``overlap'' occurred, where rotations that had already been detected were detected again.
This overlap occurred 305 times out of a total of 1,979 turns.
Notably, overlaps tend to occur when there is a prolonged change in the yaw angle, making it inappropriate to consider it as a single point.
As there are many instances of overlap where the same point is recognized as multiple turns, the missed detection rate significantly increases.

To evaluate the effectiveness of the added IF algorithm for reducing the missed detection rate, we examined its ability to detect the missed detection portion of the original PELT algorithm.
Using the same dataset, 80\% of the missed detection cases of PELT were accurately detected as turns. 
This led to a reduced missed detection rate of 8.93\%, while the false alarm rate remained at 6.97\%.

\begin{table} 
\centering
\caption{Performance comparison between two PELT-based algorithms in turn detection}
\label{tab:pelt}
\begin{center}
{\renewcommand{\arraystretch}{1.4}
 \begin{tabular}[c]{>{\centering\arraybackslash}m{0.1cm}>{\centering\arraybackslash}m{1.3cm}>{\centering\arraybackslash}m{2.0cm} >{\centering\arraybackslash}m{2.0cm}>{\centering\arraybackslash}m{2.0cm}}
 \toprule
    {} & {} & \thead{Missed detection} 
       & \thead{False alarm} \\
    \noalign{\vspace{0pt}}
\hline
 & Existing algorithm & 44.67\% & 6.97\%  \\
 \noalign{\vspace{1pt}}
 \midrule
 & Proposed algorithm & 8.93\% & 6.97\% \\
 \noalign{\vspace{1pt}}
 \bottomrule
\end{tabular}}
\end{center}
\end{table}

\section{Conclusion}

In this study, we proposed three new approaches to detect pedestrian turning motions based on smartphone IMU information. 
The limitations of existing methods in turn detection were investigated. 
Among the three proposed algorithms, the HMM-based algorithm demonstrated the best performance with a missed detection rate of 5.14\% and a false alarm rate of 2.00\%. 
This turn detection algorithm can be applied to enhance indoor map matching performance for pedestrian navigation systems.

\section*{Acknowledgment}

This work was supported in part by the Institute of Information and Communications Technology Planning and Evaluation (IITP) funded by the Korean government (KNPA) under Grant 2019-0-01291; in part by the Future Space Navigation and Satellite Research Center through the National Research Foundation of Korea (NRF) funded by the Ministry of Science and ICT (MSIT), Republic of Korea, under Grant 2022M1A3C2074404; in part by the Unmanned Vehicles Core Technology Research and Development Program through the NRF and the Unmanned Vehicle Advanced Research Center (UVARC) funded by the MSIT, Republic of Korea, under Grant 2020M3C1C1A01086407; and in part by the Korea Institute for Advancement of Technology (KIAT) grant funded by the Korea Government (MOTIE) (P0020535, The Competency Development Program for Industry Specialist).

\bibliographystyle{IEEEtran}
\bibliography{mybibfile, IUS_publications}

\vspace{12pt}

\end{document}